\documentclass[aps,twocolumn,superscriptaddress]{revtex4}
\usepackage{graphicx}
\usepackage{dcolumn}
\usepackage{bm}
\usepackage{color}
\usepackage{xcolor}
\usepackage{amsmath}
\usepackage{amssymb}
\usepackage{amsmath}
\usepackage{float}
\usepackage{natmove}
\usepackage{multirow}
\usepackage{setspace}
\usepackage{array}
\usepackage{booktabs}
\usepackage{rotating}
\usepackage[colorlinks=true,linkcolor=blue,citecolor=blue,urlcolor=blue]{hyperref}
\usepackage{physics}
\usepackage{mathrsfs}
\usepackage{bm}
\usepackage{ulem}
\newcommand{\suppinfo}{Supplemental Material~\cite{supp-info}}
\begin{document}

%
%

\title{Magnetic response and electronic states of well defined Gr/Fe/Ir(111) heterostructure}

%
%
%
\author{Claudia Cardoso}
\affiliation{Centro S3, CNR-Istituto Nanoscienze, I-41125 Modena, Italy}
\author{Giulia Avvisati}
\affiliation{Dipartimento di Fisica, Universit$\grave{a}$ di Roma ``La Sapienza'', I-00185 Roma, Italy}
\author{Pierluigi Gargiani}
\affiliation{ALBA Synchrotron Light Source, Carrer de la Llum, 2-26 08290 Barcelona (Spain)}
\author{Marco Sbroscia}
\affiliation{Dipartimento di Fisica, Universit$\grave{a}$ di Roma ``La Sapienza'', I-00185 Roma, Italy}
\author{Madan S. Jagadeesh}
\affiliation{Dipartimento di Fisica, Universit$\grave{a}$ di Roma ``La Sapienza'', I-00185 Roma, Italy}
\author{Carlo Mariani}
\affiliation{Dipartimento di Fisica, Universit$\grave{a}$ di Roma ``La Sapienza'', I-00185 Roma, Italy}
\author{Dario A. Leon}
\affiliation{Universit$\grave{a}$ di Modena e Reggio Emilia I-41125 Modena, Italy}
\affiliation{Centro S3, CNR-Istituto Nanoscienze, I-41125 Modena, Italy}

\author{Daniele Varsano}
\affiliation{Centro S3, CNR-Istituto Nanoscienze, I-41125 Modena, Italy}

\author{Andrea Ferretti}
\affiliation{Centro S3, CNR-Istituto Nanoscienze, I-41125 Modena, Italy}

\author{Maria Grazia Betti}
\affiliation{Dipartimento di Fisica, Universit$\grave{a}$ di Roma ``La Sapienza'', I-00185 Roma, Italy}

%
%
\date{\today}

%
%
\begin{abstract}
 We investigate a well-defined heterostructure constituted by magnetic Fe layers sandwiched between graphene (Gr) and Ir(111). The challenging task to avoid Fe-C solubility and Fe-Ir intermixing has been achieved with atomic controlled Fe-intercalation at moderate temperature below 500 K. Upon intercalation of a single ordered Fe layer in registry with the Ir substrate, an intermixing of the Gr bands and Fe $d$-states breaks the symmetry of the Dirac cone, with a downshift in energy of the apex by about 3 eV, and well-localized Fe intermixed states induced in the energy region just below the Fermi level. First principles electronic structure calculations show a large spin splitting of the Fe states, resulting in a majority spin channel almost fully occupied and strongly hybridized with Gr $\pi$-states.
 X-Ray Magnetic Circular Dichroism on the Gr/Fe/Ir(111) heterostructure reveals an ordered spin configuration with a ferromagnetic response of Fe layer(s), with enhanced spin- and orbital-configurations with respect to the bcc-Fe bulk values. The magnetization switches from a perpendicular easy magnetization axis when the Fe single layer is lattice matched with the Ir(111) surface, to a parallel one when the Fe thin film is almost commensurate with graphene. 
\end{abstract}

\maketitle

%
\section{Introduction}
%
Graphene can be a very promising spin channel material owing to its long spin-diffusion lengths of several micrometers \cite{Tombros2007}, gate-tunable carrier concentration, and high electronic mobility \cite{Han2014}. Graphene coupled with ferromagnetic systems can open new perspectives when efficient injection of spin-polarized electrons can be achieved, as observed for a graphene membrane on Co(0001) \cite{Usachov_2015}. It is well known that nearly flat epitaxial graphene of high structural quality can be formed on several magnetic 3$d$ metal substrates, like the lattice-matched Ni(111)~\cite{PhysRevB.84.205431, Pacile2013, Varykhalov_PRX_2012, massimi2014} and  Co(0001)~\cite{Decker_PRB_2013, Usachov_2015, Pacile_PRB_2014, Vita_PRB_2014} surfaces. It has been shown that a tiny magnetic moment arises on the carbon atoms, induced by the strong hybridization of graphene $\pi$-orbitals with Ni or Co 3$d$-states ~\cite{Usachov_2015, Weser2010, Decker_JPhys_2014, Decker_PRB_2013}. Furthermore, graphene grown on metals protects highly reactive magnetic surfaces and stabilizes them against oxidation \cite{Coraux_JPCL_2012,Cattelan_NanoScale_2015,lodesani2019graphene}.  
Whereas recently a large research effort has been dedicated to investigate Gr-Ni and Gr-Co heterostructures, only a few experimental results for graphene grown on Fe surfaces are available \cite{Vinogradov_2012, Vinogradov_2017, Varykhalov2015}, though iron is the most widespread transition metal, and the technology of passivated Fe films with a graphene membrane can be appealing for several reasons. Among them, the considerable price reduction in comparison with other transition metal substrates and, most importantly, its strong magnetic response. 

The main hurdle for the formation of graphene on top of Fe surfaces is related to the rich Fe-C phase diagram~\cite{Okamoto1992}. In fact, the high carbon diffusion coefficient and solubility in iron are detrimental for chemical vapor deposition processes, where a high annealing temperature is necessary to ensure high quality graphene on top of the metal surfaces. Thus, the epitaxial growth of a graphene membrane on a Fe film/single crystal is made difficult because of the formation of iron carbide, which is thermodynamically favored \cite{Okamoto1992}. The epitaxial growth of graphene on Fe is also prevented because none of the bcc-Fe facets is lattice-matched with graphene, at variance with close-packed surfaces of other 3$d$ metals, like Co and Ni. Recently, a detailed structural study has demonstrated that the Fe(110) face, with its distorted hexagonal symmetry, can be a good candidate, because of a partial match with the graphene lattice \cite{Vinogradov_2012, Vinogradov_2017}. In that study, the substrate temperature was kept below 675 K  to guarantee the formation of a graphene membrane on iron reducing the presence of iron carbides or segregation of diluted carbon from iron. 

A more challenging strategy has been proposed exploiting Fe intercalation beneath graphene \cite{Vlaic_JPCL_2018} as a viable route to overcome the hurdles to realize a direct growth on Fe surfaces. Intercalation of metals below Gr has proven to efficiently lead to the formation of atomically smooth metallic layer(s) \cite{Decker_PRB_2013,Pacile_PRB_2014,Vita_PRB_2014,Decker_JPhys_2014, Avvisati_ASS_2020}, in a layer-by-layer growth mode \cite{Gargiani_NatCom_2017}, where the graphene cover acts as a protective membrane of the confined epitaxial metallic layer(s) \cite{Coraux_JPCL_2012,Cattelan_NanoScale_2015}. Recently, a scanning tunnelling microscopy (STM) experiment demonstrates that successful Fe intercalation under graphene grown on Ir(111) can be obtained with the substrate temperature in the range between 450 K and 550 K \cite{Bazarnik_SurfSci_2015}, giving rise to a smooth Fe layer pseudomorphic with Ir(111) and a highly corrugated graphene membrane on top.   

In this work, following this successful strategy for Fe growth, we have realized well-defined smooth Fe layers intercalated between Gr and Ir(111), preventing any alloying and Fe-C intermixing. By means of a combined experimental and theoretical approach, we gain a  detailed insight into the physical properties of the iron-intercalated Gr/Ir(111) heterostructure. The sandwiched Fe layers present a redistribution of the minority and majority electronic density of states triggered by the spatial confinement and by the peculiar strained structural configurations, as predicted by first principles spin resolved electronic structure calculations and experimentally confirmed by angular resolved photoemission and X-Ray magnetic dichroism.

\section{Methods}
\label{sec:methods}
Angular resolved photoelectron spectroscopy (ARPES) and low-energy electron-diffraction (LEED) experiments were carried out at the Nanostructure at Surfaces laboratory, Department of Physics, Sapienza University \cite{LOTUS}. The ARPES apparatus is equipped with a Scienta SE200 multi-channel-plate electron analyzer and a monochromatic Gammadata VUV 5000 microwave He source, with main lines at 20.218 eV (He I$_{\alpha}$) and 40.814 eV (He II$_{\alpha}$) photon energies. The ARPES apparatus was set for an experimental energy and angular resolution of 16 meV and 0.18$^{\circ}$, respectively. 

The X-ray Magnetic Circular Dichroism (XMCD) measurements were performed at the BOREAS beamline of the ALBA synchrotron radiation facility~\cite{boreas}. Data were taken in the total electron yield (TEY) mode, by measuring the drain current with respect to a clean gold grid, used for photon flux normalization. We used two different geometries, grazing incidence (GI) with 70$^{\circ}$ incidence angle and normal incidence (NI), so to probe the magnetic signal along the easy and hard magnetic axes; further details are available in \cite{Avvisati_NanoLett_2018}. 

In all laboratories, the Ir(111) surface was cleaned by cycles of ion sputtering (Ar$^{+}$, 1.5-2.0 keV) and annealing (1 minute at temperatures higher than 1300 K). The quality of the obtained surface is confirmed by the sharp LEED pattern.
The Gr layer was prepared by thermal decomposition of ethylene, by exposing the clean Ir surface to 5$\times$10$^{-8}$-2$\times$10$^{-7}$ mbar of C$_2$H$_4$ and annealing at around 1300-1320 K. Metallic Fe was deposited, at 0.3 \AA/min on the Gr/Ir(111) surface kept at about 500 K, in order to ensure a high quality of Fe layer(s) intercalated beneath Gr and to avoid any intermixing, that occurs at higher annealing temperature (600-800K) \cite{Brede2016}. 
The C $1s$, Ir $4f$ and Fe $2p_{3/2}$ core-levels as measured at the Superesca beamline of the Elettra synchrotron radiation facility, are presented in the \suppinfo.  

The completion of a single layer of Fe can be identified by following the evolution of the $\pi$ states of Gr in the ARPES valence band, as discussed in detail in the next sections. At the  BOREAS beamline, the Fe density in the intercalated film was evaluated via its jump-edge ratio, previously calibrated with Auger electron spectroscopy, as reported in the Supporting Information of Ref.~\cite{Gargiani_NatCom_2017}.

Density functional theory (DFT) simulations were carried out using the \textsc{Quantum ESPRESSO} package~\cite{Giannozzi2009,Giannozzi2017} where wavefunctions are expanded in plane-waves and pseudopotentials are used to account for the electron-ion interaction. We used the local density approximation (LDA) for the exchange-correlation potential, according to the Perdew-Zunger parametrization~\cite{Perdew_PRB_1981}. 
Similarly to our previous works~\cite{Avvisati_NanoLett_2018,  CalloniJCP2020}, we simulated the Gr/1ML-Fe/Ir(111) interface considering the complete moir\'e induced periodicity by using a 9$\times$9 supercell of Ir(111), corresponding to a 10$\times$10 supercell of pristine Gr. The lattice parameters were obtained by relaxing  Ir bulk at the LDA level using ultrasoft pseudopotentials (USPP), resulting in a Ir--Ir bond distance of 2.7048~\AA{} (corresponding to a hexagonal cell of 46.001 Bohr radius for the moir\'e structure). In all the calculation we included four metallic layers (3 Ir plus one Fe layer or 4 Ir layers). In order to make the two sides of the slab inequivalent we added a layer of H atoms in one of the two sides. Atomic positions were then fully relaxed (except for the two bottom Ir layers and the H saturation layer) until ionic forces were smaller than 0.001 Ry/Bohr. For all the self-consistent calculations we used a 2$\times$2 grid of $\mathbf{k}$-points, ultra soft pseudopotentials to model the electron-ion interaction and a kinetic energy cutoff of 30 and 300 Ry to represent Kohn-Sham wavefunctions and  density, respectively.

\begin{figure*}
\centering
\includegraphics[clip,width=0.9\textwidth]{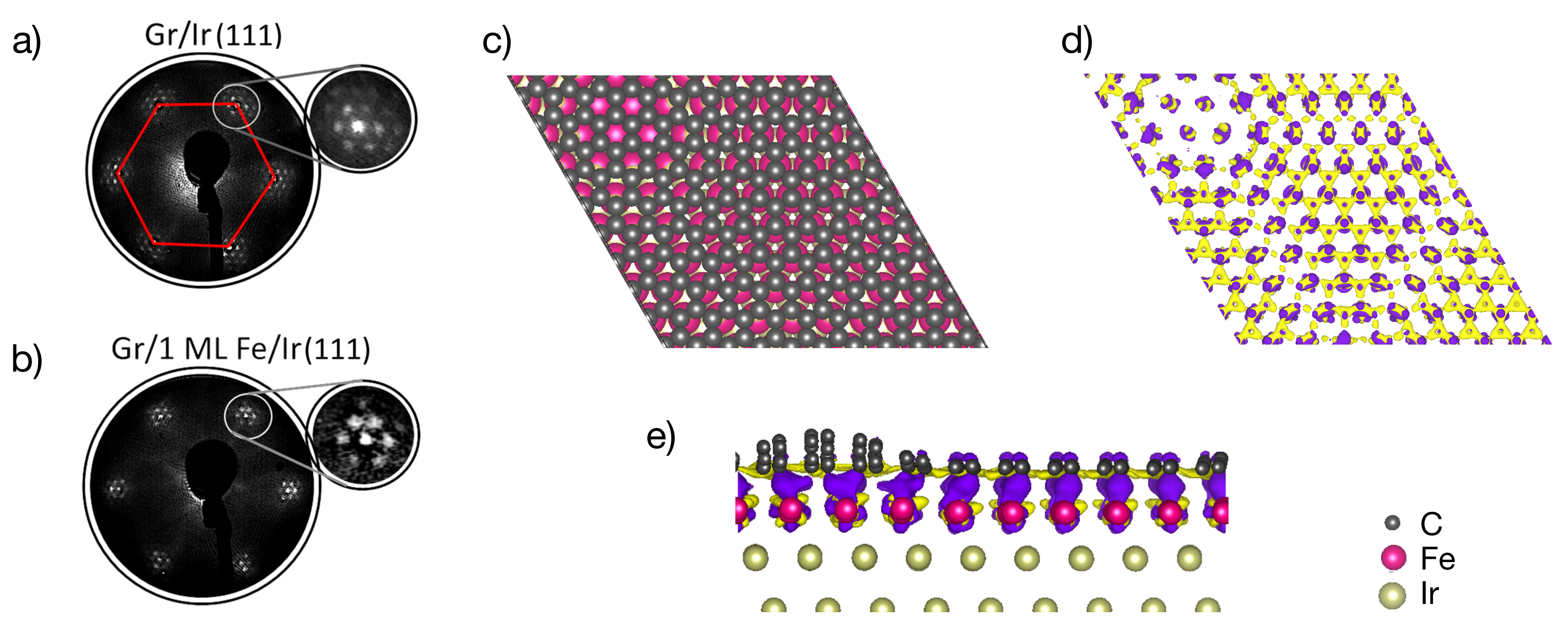}
\caption{\label{fig:growth} 
Low energy electron diffraction (LEED) patterns  (electron beam energy 140 eV) for (a) Gr/Ir(111) and (b) Gr/1 ML Fe/Ir(111); (c) atomic structure as deduced by DFT, top view of the  moir\'e pattern of Gr/Fe/Ir(111) with C atoms represented in gray, Fe in red and Ir in cream; (d-e) charge difference with respect to the free standing Gr and the pristine Fe/Ir(111) slab, computed with DFT: (d) top view of the charge difference isosurfaces (positive shown in purple, negative in yellow) and (e) side view of the Gr/Fe/Ir(111) structure showing the graphene corrugation and charge difference isosurfaces, using the same colors described for the previous panels.
}
\end{figure*}
Since LDA is known to underestimate the values of the orbital magnetic moments in transition metals~\cite{Truhlar09}, we have adopted a DFT+U scheme~\cite{cococcioni2005linear}, with a Hubbard U parameter of 2~eV.
In order to compare calculated band structures with ARPES data, we applied an unfolding procedure~\cite{Popescu2012,unfold-x} to the computed bands of Gr/Fe/Ir(111) and Gr/Ir(111). With this procedure, the band structure computed for the 10$\times$10 supercell is mapped into the graphene 1$\times$1  Brillouin zone by using the {\tt unfold-x} code \cite{unfold-x}. In this way we obtain an effective band structure corresponding to the graphene unit cell. 
In this picture, the $\mathbf{k}$-dispersion is broadened by the break of the 1$\times$1 translational symmetry induced by the 10$\times$10 moir\'e pattern.

\section{Results and Discussion}
%
\label{sec:results_intercalation}
\subsection{Intercalation and structural properties} 
\begin{figure*}
\centering
\includegraphics[clip,width=0.75\textwidth]{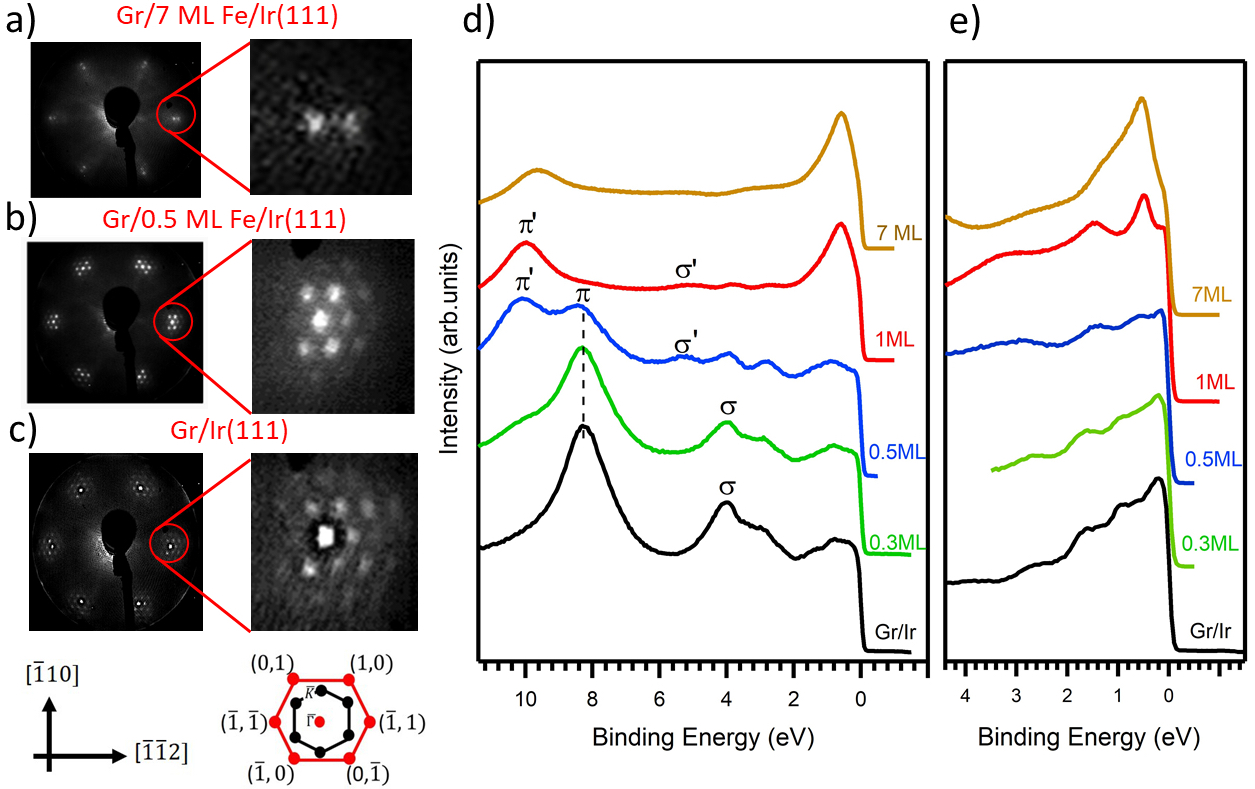}
\caption{\label{fig:arpes}
 LEED patterns  (electron beam energy  140 eV) and zoomed pattern around the (10) diffraction spot for Gr/7ML Fe/Ir(111) (a), Gr/0.5ML Fe/Ir(111) (b) and Gr/Ir(111) (c); below: sketch of the main symmetry directions and of the diffraction pattern of the Ir(111) surface (red dots); the Surface Brillouin Zone (SBZ) is shown in black. Photoemission Energy Distribution curves (EDC) taken at 40.814 eV (He II$_{\alpha}$); angular integrated spectra around  $\Gamma$ (d) and $K$ (e) points of the SBZ for different thickness of the intercalated Fe layer between Gr and Ir}
\end{figure*} 

A single layer of iron, intercalated under Gr grown on Ir(111), induces a corrugation of the Gr membrane and preserves the periodicity of the  moir\'e superstructure, superimposed to the hexagonal Gr lattice, as unraveled by the LEED patterns reported in Fig.~\ref{fig:growth}(a,b). The diffraction pattern of Gr/Ir(111) (Fig.~\ref{fig:growth}a) shows bright spots in a hexagonal pattern surrounded by satellites, consistent with the moir\'e superstructure caused by the lattice mismatch between Gr and Ir \cite{NDiaye08, Scardamaglia2011}. After Fe intercalation, the pattern is only slightly attenuated (Fig.~\ref{fig:growth}b), thus the first Fe layer appears commensurate to the Ir(111) surface lattice. This evidence is corroborated by a recent work~\cite{Decker_JPhys_2014} where STM measurements show that a few monolayers (ML) of Fe intercalated under graphene arrange in registry with the hexagonal Ir(111) surface, with a corrugation of 1.3~\AA, larger than the one measured for Gr/Ir(111) \cite{PhysRevLett.107.036101}. 

We have further investigated the structural properties of this system by means of DFT at the LDA+U level. The Gr/Fe/Ir(111) heterostructure was modeled as described in Sec.~\ref{sec:methods}. The calculations confirm the large corrugation of the graphene layer (1.44~\AA), with a minimum graphene-Fe interplanar distance of 1.85~\AA\ and a maximum of 3.29~\AA\ on the crests of the hills, as shown in Fig.~\ref{fig:growth}(c). 
Previous DFT calculations~\cite{Decker_JPhys_2014} done at the PBE level including van der Waals interactions by using the DFT–D2 method~\cite{Grimme2006JCC,Decker_D2_note} found a similar scenario, with slightly larger Gr-Fe distances (2.05 and 3.33 \AA{} for valleys and hills, respectively).

The topography of the Gr/Fe/Ir(111) heterostructure is similar to the corrugated moir\'e superstructure observed for the Gr/Co/Ir(111) system \cite{Avvisati_NanoLett_2018}, but very different from the structure reported for Gr directly grown on bcc-Fe(110)~\cite{Vinogradov_2012}.
Previous STM measurements and DFT calculations performed at the LDA level, for Gr/bcc-Fe(110) point out the formation of a periodic corrugated pattern of the graphene layer parallel to the [001] direction of the substrate, consisting in a supercell of 7$\times$17 graphene hexagons with a smaller corrugation of 0.6/0.9~\AA, and only a small fraction of the C atoms considerably elevated over the Fe surface, thus making the entire graphene membrane strongly interacting with the metal substrate.

In the present case, the larger graphene corrugation modulates the Gr/Fe interaction. This is illustrated by the charge difference computed for Gr/Fe/Ir(111) shown in Fig.~\ref{fig:growth}(d,e). The excess of negative charge (yellow isosurface) is accumulated in the graphene membrane, donated by the Fe intercalated layer. The redistribution of charge is more pronounced in the valley regions and  milder  in the areas corresponding to the crests of the hills, corroborating a different strength of hybridization between graphene and Fe going from the valleys to the crests due to the increasing graphene–Fe distance. The periodicity of strongly and weakly bound regions in which covalent and van der Waals bonding dominate, respectively, induces a different balance in the charge transfer determined also by the registry with respect to the underlying Fe atoms. 

Another signature of the similarity of the registry and graphene corrugation resulting from Co and Fe intercalated layers can be deduced by comparing the C~1$s$ core level photoemission data for Gr/Fe/Ir(111) (as reported in \suppinfo) and Gr/Co/Ir(111). In fact, there is a multi-component C~1$s$ line-shape for both systems, in which the two main features are assigned to C atoms in the Gr layer weakly (on the crests) and strongly (in the valleys) bound to the Co and Fe layers \cite{Pacile_PRB_2014,Avvisati_JPCC_2017,Avvisati_ASS_2020}. In contrast, when graphene is grown directly on the bcc-Fe(110) surface, a single C 1$s$ core level component is present, explained by the presence of extended regions of graphene coupled to the Fe(110) substrate, even at the crests of the moir\'e superstructure \cite{Vinogradov_2012, Vinogradov_2017}. For the graphene membrane directly grown on Fe(110), a small feature in the C 1s core level has been detected at lower binding energy (BE), linked to the presence of Fe carbide resulting from the Fe-C solubility due to the high temperature reached during the chemical vapor deposition growth procedure \cite{Vinogradov_2012, Vinogradov_2017}. 
The similarity of the Fe intercalation process under Gr/Ir(111) with the Gr/Co/Ir(111) heterostructure and the absence of a C 1$s$ component due to the formation of iron carbide (see details in \suppinfo) demonstrate the successful strategy to intercalate Fe under graphene at low temperature, preventing any solubility into C. 

The corrugation of the Gr membrane discussed so far is attenuated when the number of Fe intercalated layers increases above a few ML, with a reduction of the brightness of the superstructure LEED spots, as reported in Fig.~\ref{fig:arpes}(a-c).
At 7 ML, we observe the formation of a
Fe film with a lattice constant smaller than Ir(111) and therefore closer to the graphene structural parameters. However, even when the strain induced by the Ir(111) surface is released, the Fe film remains slightly incommensurate with graphene, as shown in Fig. \ref{fig:arpes}(a). This is in contrast with the case of Co intercalation beneath graphene on Ir(111): when the Co film is formed, it recovers its hpc lattice constant and becomes commensurate with the graphene layer \cite{Pacile_PRB_2014,Avvisati_JPCC_2017}. Indeed, while the hcp Co(0001) surface lattice parameter  (2.50~\AA), is comparable with the 2.46~\AA{} of Gr, none of the bcc-Fe faces is lattice-matched with graphene. In fact, graphene directly grown on bcc-Fe(110) presents a distorted hexagonal symmetry with only a partial match with the  graphene lattice, as deduced by STM \cite{Vinogradov_2012}. In our case, the Fe intercalated multilayer has a novel strained structural configuration with a 8\% mismatch with respect to the graphene lattice constant (see details of the LEED spot intensity analysis in the \suppinfo). 

Importantly, the magnetic response of Fe films is extremely sensitive to tiny variations of the structural configuration \cite{Moruzzi1986}, and the strained lattice of the Fe film intercalated under graphene can give rise to an altered distribution of the electronic majority and minority states and to different spin and orbital configurations with respect to the bulk reference \cite{Moulas_PRB_2008}.

\begin{figure}
\centering
%
%
{\footnotesize $\quad$ Gr/Ir(111) $\qquad \qquad \qquad \quad\quad$ Gr/Fe/Ir(111) }
\includegraphics[clip,width=0.48\textwidth]{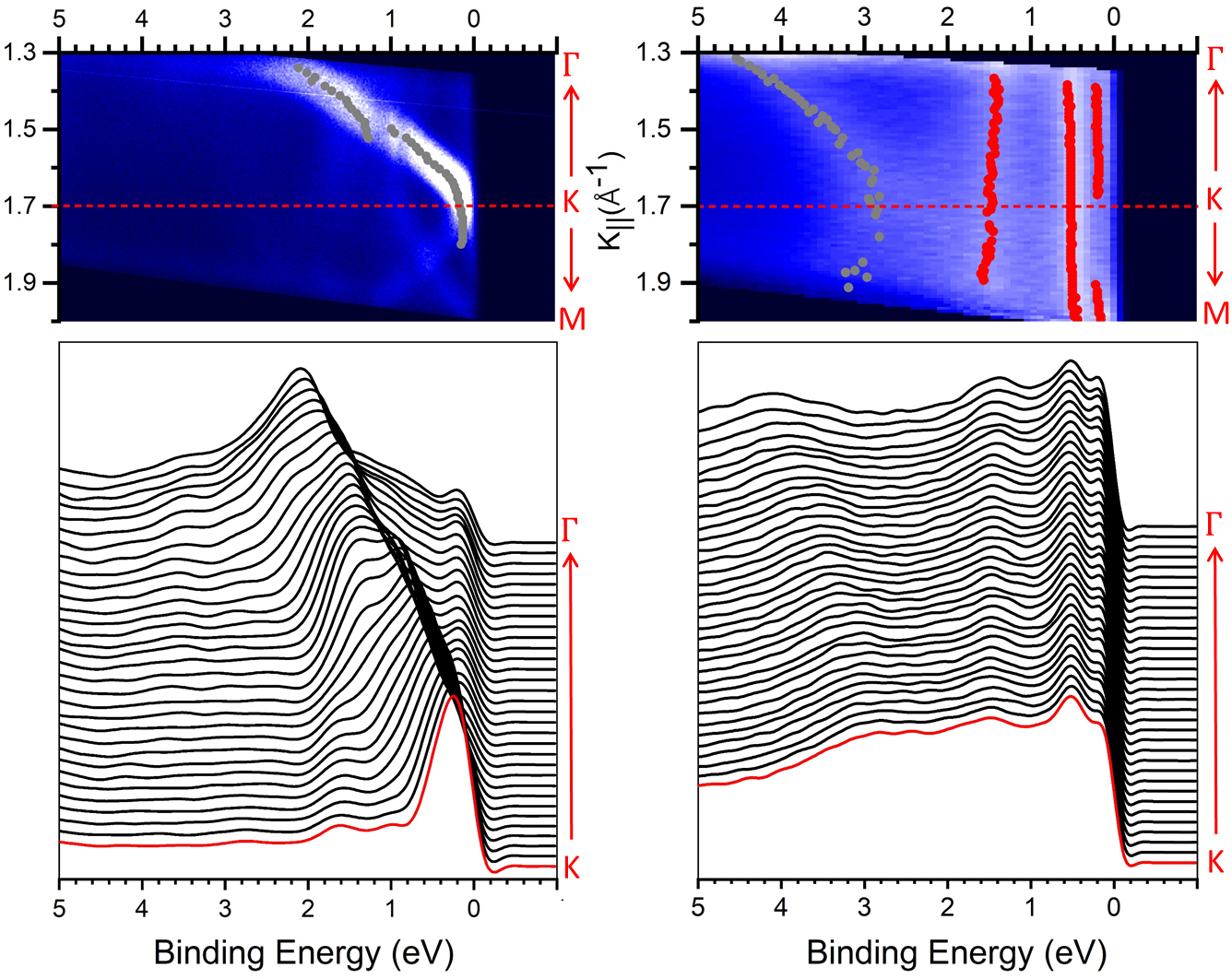}
\caption{\label{fig:ARPES-K}
 Angular resolved photoelectron spectroscopy h$\nu$=40.814 eV  spectra of Gr/Ir(111) (left panels) and Gr/1 ML Fe/Ir(111) (right panels), around the K point of the SBZ. Electronic band structure vs. k$_{//}$ as intensity plot (upper panels); spectral density plotted as sequential EDC curves from the K point towards $\Gamma$ (bottom panels).}
\end{figure}

\subsection{Electronic structure}

\begin{figure*}
{\footnotesize \hspace{0.0cm} Gr/Ir(111) \hspace{5.0cm} Gr/Fe/Ir(111) }
%
\includegraphics[clip,width=0.85\textwidth]{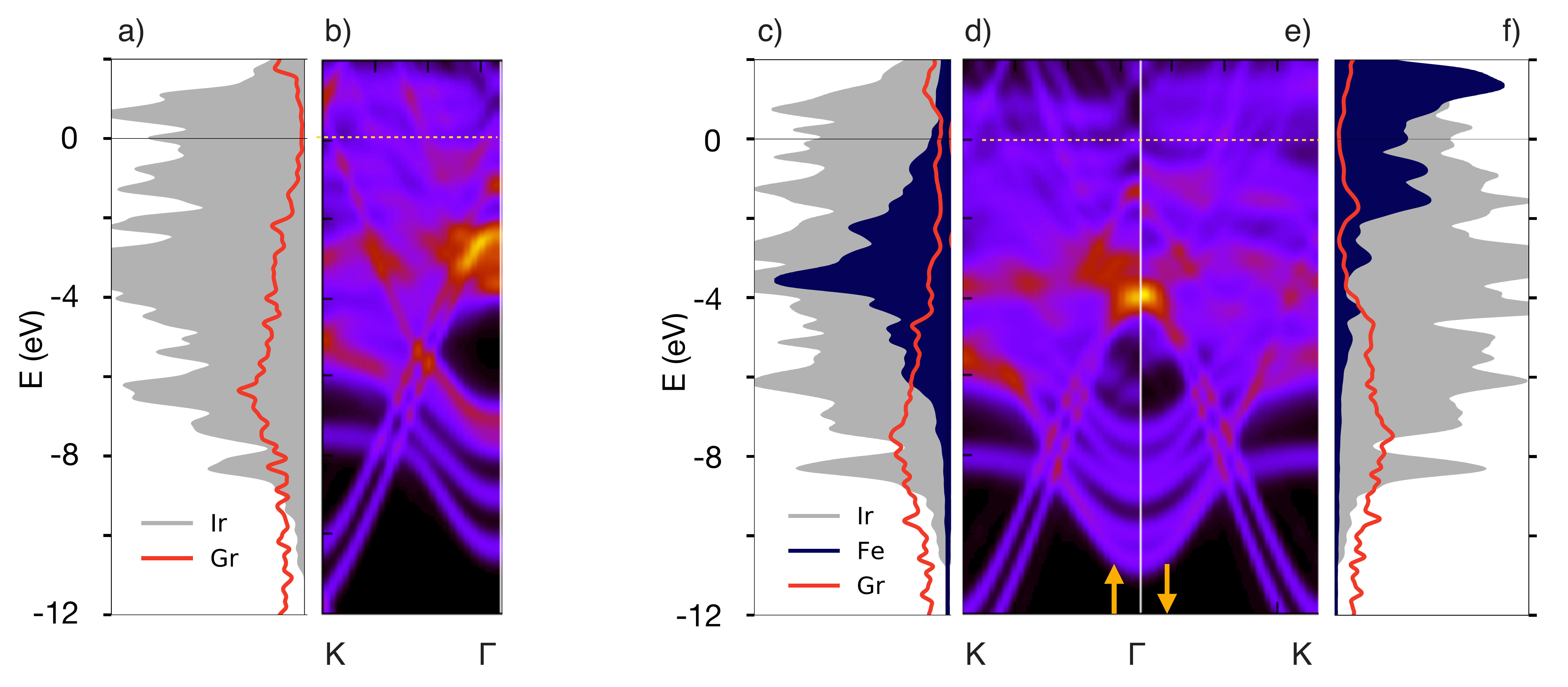}
\caption{\label{fig:VB} 
Band Structure and projected Density of States (pDOS) computed within DFT for Gr/Ir(111) and Gr/1 ML Fe/ Ir(111): a) Gr/Ir(111) DOS projected on C and Ir atomic orbitals and b) Gr/Ir(111) band structure, unfolded on the 1$\times$1 graphene unit cell as described in the main text; c) Gr/Fe/Ir(111) majority spin DOS projected on C, Fe and Ir atomic orbitals; d) majority  and e) minority spin band structure, unfolded on the 1$\times$1 graphene unit cell; f) Gr/Fe/Ir(111) minority spin pDOS.
}   
\end{figure*}

A deeper insight into the interaction between Gr and the Fe intercalated layer(s) can be unveiled by the electronic spectral density deduced by angular resolved photoelectron spectroscopy of the Gr/Fe/Ir(111) heterostructure, compared with ab initio theoretical predictions of the electronic density of states (DOS) and band structure calculated by DFT.

{\it Experimental data.}
The photoemission spectral density as a function of Fe intercalation thickness is reported in Fig.~\ref{fig:arpes} (d) and (e), at the $\Gamma$ and K points of the surface Brillouin zone (SBZ),  respectively. The Gr/Ir(111) valence band at the $\Gamma$ and K points presents the expected electronic spectral density \cite{PhysRevLett.102.056808,PhysRevB.84.075427} whereas the intercalated system shows three new features when compared  with the bare Gr/Ir(111) : ($i$) the appearance of a peak at 10~eV and the disappearance of the one related to the $\pi$- bands of bare Gr/Ir(111) at 8 eV at the $\Gamma$ point; ($ii$) the quenching of the intensity from the Ir surface states close to the Fermi level; and ($iii$) the emergence of extra spectral density in the low binding energy region.

Concerning ($i$), the intensity of the peak at 8~eV BE at $\Gamma$, corresponding to the bottom of the $\pi$-band for bare Gr/Ir(111), decreases with increasing Fe thickness. The peak emerging at 10.0~eV BE can be associated to the shifted $\pi$-band, due to the interaction of Gr with the intercalated Fe layer. An interacting $\pi$-band has also been observed for Co and Ni layers sandwiched between Gr and Ir \cite{Pacile_PRB_2014, Vita_PRB_2014, Pacile2013, massimi2014}. The presence of two features associated with the $\pi$-bands, with opposite intensity behavior before the completion of the first Fe ML, demonstrates the coexistence of bare Gr/Ir(111) regions and regions of intercalated Fe atoms, up to the formation of a smooth Fe single layer, when the peak at 8~eV (bare Gr/Ir(111)) is eventually quenched. 

Furthermore, ($ii$) the progressive intensity lowering of the Ir(111) surface states in the energy region 0-3 eV BE, upon Fe intercalation, without any energy shift, suggests the absence of Fe-Ir intermixing observed at higher intercalation temperatures~\cite{Bazarnik_SurfSci_2015, Brede2016}. 
The emergence of a spectral density ($iii$) at low binding energy close to the Fermi level can be ascribed to electronic states mainly localized in the Fe layer, since their intensity definitely raises at increasing Fe thickness, as can be clearly observed in the spectral density of 7 ML Fe intercalated under graphene (orange curve in Fig.~\ref{fig:arpes}e).

A clearer assignment of the electronic states due to the Fe-Gr interaction process can be derived from the electronic state dispersion. The ARPES data around the $K$ point of the SBZ for Gr/Ir(111) and Gr/1 ML Fe/Ir(111), are shown in Fig.~\ref{fig:ARPES-K} (left and right panels, respectively).
Considering the energy region closer to the Fermi level, three localized Fe states can be identified at about 0.2~eV, 0.5~eV and 1.5~eV BE, and found to be only slightly dispersing over the whole SBZ (Fig. \ref{fig:ARPES-K} right panel), as expected for a confined Fe film. Inspecting the spectra at higher binding energies, other Fe-related states are observed in the energy region of 3-4 eV BE, resonant with the graphene $\pi$-band states.
Besides the presence of these localized Fe states close to the Fermi level, the most evident consequence of Fe intercalation under graphene is the downshift of the Dirac cone at the $K$ point of the SBZ, similarly to the case of Gr/Co/Ir(111)~\cite{Pacile_PRB_2014}.
In fact, the $\pi$-band, as deduced by the EDCs (Fig. \ref{fig:ARPES-K}), is broadened and less defined after intercalation and there is a coexistence of other Fe related states in this energy region (3-4 eV). 
The position of the Dirac cone vertex (at K) is downshifted by 3~eV, while the bottom of the $\pi$ band is less shifted at $\Gamma$. It is worth noting that the Dirac cone is shifted by 2~eV towards higher binding energy also for graphene directly grown on Fe(110)~\cite{Varykhalov2015}, suggesting a weaker interaction in that case.


{\it Theory.}
The measured ARPES data are complemented by DFT calculations, as shown in Fig.~\ref{fig:VB}, where the electronic structures computed for Gr/Ir(111) and Gr/Fe/Ir(111) are unfolded and mapped into the 1$\times$1 graphene Brillouin zone along the $\Gamma$ -- $K$ direction, as described in Sec.~\ref{sec:methods}. Projected DOS (and projected bands in \suppinfo) are also provided to complement the band information.
The bands obtained for Gr/Ir(111) are in good agreement with existing literature, as from e.g. Ref.~\cite{Voloshina2015JPCM}. Concerning the Fe intercalated system, as also observed in the experiments, the Gr $\pi$- and $\sigma$-bands are shifted to higher binding energies by the effect of Fe intercalation ($i$).
While for Gr/Ir(111) the $\pi$ bands are clearly recognizable, in the case of Gr/Fe/Ir(111) they are strongly hybridized with Fe states above -6~eV and the upper part of the cone is identifiable only for the minority spin states. The bottom of the $\pi$ bands at $\Gamma$ moves from about -7.5~eV in Gr/Ir(111) to about -10.0 eV in Gr/Fe/Ir(111) (see Fig.~S4 in \suppinfo), with an overall downshift of $\sim$2.5 eV, in quite good agreement with the experimental findings.
Similarly, the $\sigma$ bands of graphene undergo a downshift by about -1.5 eV at both $\Gamma$ (going from -2.5 to -4 eV) and $K$ (from -10.5 eV to -12 eV). 
While the experimental offset of these features is found at larger binding energies (namely 4.0 eV and 5.5 eV as marked by $\sigma$ and $\sigma'$ in Fig.~\ref{fig:arpes}d), the downshift is quite consistent with the calculations~\cite{note_ARPES_DFT}. 

In order to identify Fe-related states, we consider the projected DOS, shown in Fig.~\ref{fig:VB}(c,f), where a strong spin splitting of the Fe components is observed (see discussion below).
For instance, a large peak of the Fe pDOS is found in the energy region of 3-4~eV BE, in good agreement with the experimental data, and mainly located in the majority spin channel. Such peak then overlaps with the $\pi$ bands of Gr disrupting the Dirac cone for majority states. In contrast, the Dirac cone is still faintly visible in the minority spin bands at about 2.5-3 eV.
In both spin channels the pDOS reveals overlapping peaks due to Fe and graphene states, suggesting that the shadowing of the vertex of the Dirac cone observed in the experimental photoemission data is clearly induced by the hybridization of the graphene $\pi$-states with the Fe $d$ majority states. 
Furthermore, the spin resolved DOS in the energy region closer to the Fermi level shows that the main Fe-related peak of the minority spin states lies above the Fermi level, with 
 smaller peaks between -2~eV and the Fermi energy,
 in good agreement with the experimental observation ($iii$). 
 Overall, this picture is further confirmed by the projected unfolded bands provided in Fig.~S4 of \suppinfo.

\subsection{Magnetic properties and spin and orbital configuration of the Fe layer}

\begin{figure*}
\centering
\includegraphics[clip,width=0.75\textwidth]{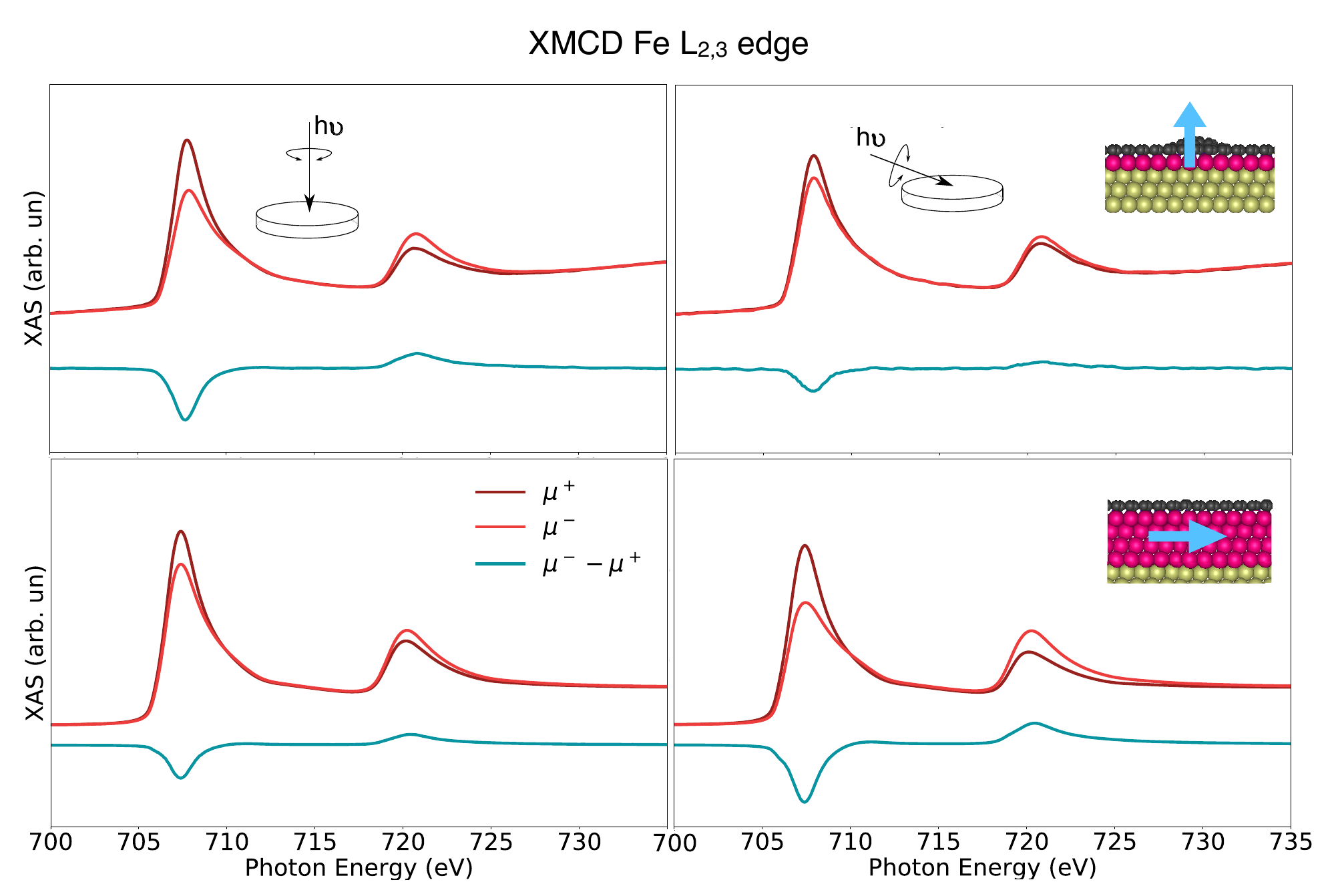}
\caption{\label{fig:XMCD} 
XMCD spectra of Fe L\textsubscript{2,3} absorption edges for Gr/1.4 ML Fe/Ir(111) (top) and Gr/7ML Fe/Ir(111) (bottom) acquired in remanence at RT in normal (left) and grazing (right) incidence geometries (see sketches for the experimental geometry). The easy magnetization direction switching from perpendicular (top) to parallel (bottom) to the surface plane is indicated by the blue arrows.}   
\end{figure*}


In contrast with Fe grown on Ir(111), for which the loss of inversion symmetry at the interface of the magnetic layer and substrate stabilizes different skyrmion lattices depending on the Fe/Ir(111) stacking~ \cite{Heinze_NatPhys_2011,vonBergmann2007,vonBergmann2006,vonBergmann2015,Grenz_PRL_2017}, when a Fe monolayer is intercalated between Gr and Ir(111), the heterostructure exhibits a ferromagnetic order with an out-of-plane easy magnetization axis~\cite{Decker_JPhys_2014}. For this reason, and based on the experimental evidence described below, in the present DFT calculations we have considered only collinear magnetic configurations, excluding in this way skyrmions or spin spiral textures.

The pDOS computed for the Gr/Fe/Ir(111) heterostructure shows a large spin split of the Fe states, with the maxima of the two spin pDOS almost 5~eV apart. In particular, the majority spin states are almost fully occupied while the largest peak of the minority spin states is empty, which results in an average Fe-magnetic moment of 2.2~$\mu_B/$atom. 
The computed ground state magnetic configuration is ferromagnetic even if the Fe magnetic moment shows a modulation over different sites ranging from 2.0 to 2.7~$\mu_B/$atom, in good agreement with previous calculation~\cite{Brede2014}. The modulations is determined by the graphene layer, with the Fe atoms located below the graphene crests, i.e. with larger Fe-C distances, having the largest values. Concerning graphene, the C magnetic moments are non-zero but quite small (at the LDA+U level ranging from -0.02 to 0.01~$\mu_B/$atom),
similarly to what reported for Gr/Co/Ir(111)~\cite{Vita_PRB_2014,Avvisati_NanoLett_2018}, with a distance- and sublattice-dependent modulation.

Overall, we find that the Fe monolayer displays a clear spin splitting, as an effect of the band narrowing and subsequent increased number of majority electrons. This is confirmed by the strong localization of the Fe-induced electronic states observed in the photoemission spectral density.
It is interesting to point out that a different scenario is found for
the Gr/Co/Ir(111) interface~\cite{Vita_PRB_2014} (see \suppinfo 
for a detailed comparison with Gr/Co/Ir(111) pDOS) where Co spectral weight in the minority channel is shifted at lower energies leading to a spin split significantly smaller than in the present case.


The strong ferromagnetic behavior is confirmed by the X ray absorption spectra and the XMCD at the L\textsubscript{2,3} edges for 1.4 ML and 7 ML of Fe sandwiched between Gr and Ir(111), shown in Fig.~\ref{fig:XMCD}. XAS are acquired in remanence conditions with circularly polarized radiation and with photons impinging normal (left panel) and grazing (right panel) to the surface. The XMCD spectra are obtained as the difference between the absorption edges acquired with left- and right-circularly polarized radiation. The higher dichroic response with photon impinging at normal incidence (left panel) reveals an out-of-plane magnetic anisotropy of the Fe layer when it is pseudomorphic to the Ir(111) surface, with a stretched Fe-Fe distance with respect the bulk bcc(110) or (111) surfaces. 
The higher dichroic response for Fe L\textsubscript{2,3} XMCD at grazing incidence (right panel) for the thicker intercalated Fe film, unveils a switch of the easy magnetization axis from perpendicular to the surface to in-plane. This is consistent with the magnetic response of heterostructures of single Co or Fe$_x$Co$_{1-x}$ layers on Ir(111) covered with a Gr membrane \cite{Avvisati_ASS_2020, Avvisati_NanoLett_2018}. 

The spin and the orbital moments at the Fe site, as deduced via XMCD sum rules, show a doubled $L/S_{\text{eff}}$ ratio ($S_{\text{eff}}=S + 7D$ and D is the dipolar moment \cite{Thole_PRL_1992,Carra_PRL_1993}), with respect to the bulk element (0.11$\pm0.01$ for a Fe single layer and 0.09 $\pm0.01$ for the 7 ML). These experimental results, related only to the intensity of the dichroic signal and independent on the number of $3d$ holes, suggest an enhanced magnetic response due to the redistribution of the majority spin states in the conduction Fe bands. The evaluated average total moment is 2.1$\pm0.2$ $\mu_B/\text{atom}$ for 7 ML of Fe intercalated under Gr, in fair agreement with 2.2~$\mu_B/$atom (spin moment), as deduced by the DFT predictions. The orbital and spin moment for a single layer of Fe, where interface effects can play a role,  shows a comparable magnetic response, as deduced by the similar dichroic signals reported in Fig.~\ref{fig:XMCD}.  As mentioned above,  the pseudomorphic  hexagonal Fe single layer grown directly on Ir(111) presents a complex magnetic structure with skyrmionic spin spiral textures stabilized by the $3d–5d$ hybridization between Fe and Ir.~\cite{vonBergmann2006, vonBergmann2007,vonBergmann2015,Grenz_PRL_2017}. 
On the other hand, other magnetic materials like Europium have been successfully intercalated underneath graphene on the same Ir(111) surface \cite{doi:10.1021/nl402797j} and, depending on its coverage, Eu displays either a paramagnetic or a ferromagnetic behavior \cite{PhysRevB.90.235437}.
The clear magnetic dichroism response of the single Fe layer intercalated under graphene suggests a different spin configuration. Furthermore, this confined single Fe intercalated layer, with stretched Fe-Fe distances, presents spin and orbital configurations similar to those of the thin Fe intercalated film, where the influence of the Fe-Ir interface is completely released.

\section{Conclusions}
The strategy to sandwich the Fe layer beneath the graphene membrane with an intercalation process at moderate temperature (500 K), prevents any alloying, and the absence of any hallmark of Fe-Ir and Fe-C intermixing proves the formation of a well-defined homogeneous Fe layer in registry with the Ir(111) surface. Such single layer of Fe, protected by the graphene membrane, induces a downshift in energy and a symmetry breaking of the Dirac cone due to the interaction between Gr and Fe majority states resonant in the energy region of the vertex of the cone. The redistribution of the spin resolved Fe pDOS  with a narrowing of the Fe bands and a larger spin splitting between majority (fully occupied) and minority states (almost empty) of the Fe states associated with increased total Fe magnetic moment influences the magnetic response of the Fe intercalated layer(s). 

In contrast to the case of direct growth of Fe on Ir(111), where, despite the large Fe magnetic moment, a non-collinear magnetic order has been observed \cite{Heinze_NatPhys_2011,vonBergmann2007,vonBergmann2006,vonBergmann2015}, the Fe layer with graphene on top reveals a ferromagnetic long range order with spin and orbital moments that are higher than the ones found for the bulk phases. The graphene top layer acts not only as a protective membrane, but also allows for a stable ferromagnetic configuration, counteracting the effect of Ir substrate. The concomitant dimensionality reduction with a narrowing of the $d$ bands and a reduced superimposition between the spin-split majority and minority Fe bands, further contribute to the transition of a single Fe layer (or few layers of Fe), from weak to strong ferromagnet, when intercalated beneath graphene.  These 3$d$ confined layers protected by a graphene membrane with a novel structural configuration with respect to the bulk lattice arrangement offer a powerful playground to tune their electronic structure and magnetic state for magnetic/spintronic devices.

\section{Acknowledgements}
  The authors thank the experimental assistance of BOREAS beamline staff at ALBA, of SuperESCA beamline staff at Elettra, and of Alessio Vegliante in LOTUS lab in Rome. G.A. and MG.B. acknowledge support from translational access CalipsoPlus funding.  C.C., D.V., A.F., D.A.L. acknowledge partially support from the MaX -- MAterials design at the eXascale -- Centre of Excellence, funded by the European Union program H2020-INFRAEDI-2018-1 (Grant No. 824143). They also thank SUPER (Supercomputing Unified Platform -- Emilia-Romagna) from Emilia-Romagna POR-FESR 2014-2020 regional funds.
  The work was partially supported by PRIN FERMAT (2017KFY7XF) from Italian Ministry MIUR and by Sapienza Ateneo funds. Computational time on the Marconi100 machine at CINECA was provided by the Italian ISCRA program. 


%
%


\begin{thebibliography}{10}
\bibitem{Decker_D2_note}
Note that the DFT-D2 correction for van der Walls interactions has been
  reparametrized in these works.
\bibitem{note_ARPES_DFT}
Note that comparing ARPES data with KS-DFT calculations absolute quantitative
  agreement in the position of the features cannot be expected.
\end{thebibliography}
\end{document}


%
%
\title{Magnetic response and electronic states of well defined Gr/Fe/Ir(111) heterostructure
  \\Supplemental Material}

%

%
\author{Claudia Cardoso}
\affiliation{Centro S3, CNR-Istituto Nanoscienze, I-41125 Modena, Italy}

\author{Giulia Avvisati}
\affiliation{Dipartimento di Fisica, Universit$\grave{a}$ di Roma ``La Sapienza'', I-00185 Roma, Italy}

\author{Pierluigi Gargiani}
\affiliation{ALBA Synchrotron Light Source, Carrer de la Llum, 2-26 08290 Barcelona (Spain)}
\author{Marco Sbroscia}
\affiliation{Dipartimento di Fisica, Universit$\grave{a}$ di Roma ``La Sapienza'', I-00185 Roma, Italy}

\author{Madan S. Jagadeesh}
\affiliation{Dipartimento di Fisica, Universit$\grave{a}$ di Roma ``La Sapienza'', I-00185 Roma, Italy}
\author{Carlo Mariani}
\affiliation{Dipartimento di Fisica, Universit$\grave{a}$ di Roma ``La Sapienza'', I-00185 Roma, Italy}

\author{Dario A. Leon}
\affiliation{Universit$\grave{a}$ di Modena e Reggio Emilia I-41125 Modena, Italy}
\affiliation{Centro S3, CNR-Istituto Nanoscienze, I-41125 Modena, Italy}

\author{Daniele Varsano}
\affiliation{Centro S3, CNR-Istituto Nanoscienze, I-41125 Modena, Italy}

\author{Andrea Ferretti}
\affiliation{Centro S3, CNR-Istituto Nanoscienze, I-41125 Modena, Italy}

\author{Maria Grazia Betti}
\affiliation{Dipartimento di Fisica, Universit$\grave{a}$ di Roma ``La Sapienza'', I-00185 Roma, Italy}

%
%
\keywords{intercalation;  photoemission; graphene; moir\'e structure; Ir(111)}
\pacs{}
\date{\today}

\maketitle

\section{Core levels}
\begin{figure}
\centering
\includegraphics[clip,width=0.40\textwidth]{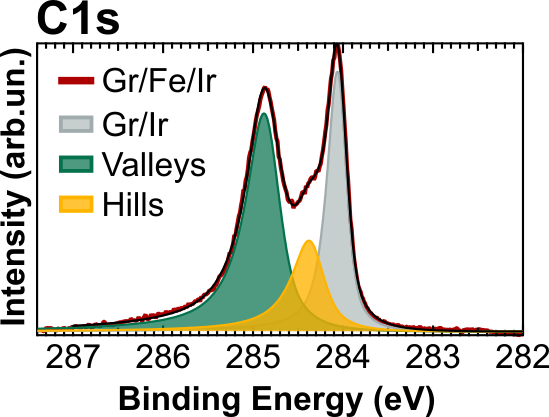}
\caption{\label{fig:C1s} C 1\textit{s} core level, taken at $h\nu$=400 eV, of Gr/0.8ML-Fe/Ir(111). Experimental data (red dots), fitting curve (red line) and single fitting components (colored curves). The fitting procedure unveils three components: the C 1s related to the Gr/Ir(111) (gray curve), the C 1s component of the C atoms in the valleys (green curve) and in the hills (yellow curve) of the moir\'e superstructure. }
\end{figure} 

\begin{figure}
\centering
\includegraphics[clip,width=0.40\textwidth]{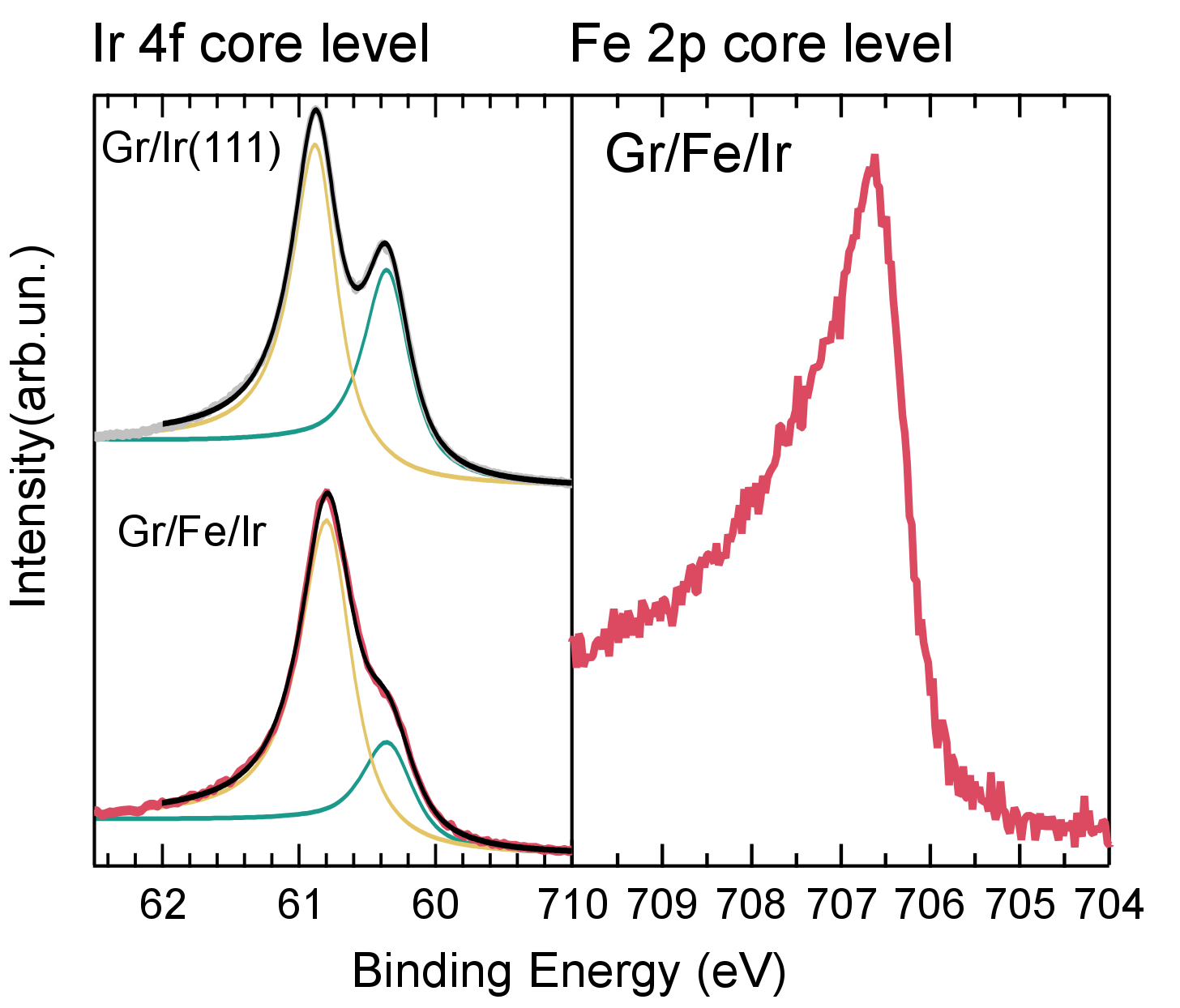}
\caption{\label{fig:Fe2p} Left panel: Ir 4f$_{7/2}$ core levels, taken at $h\nu$=400~eV, of Gr/Ir(111) (top gray curve) and Gr/0.8 ML Fe/Ir(111) after intercalation (bottom red curve). The fitting curves are drawn as black lines, and the single fitting components are reported in yellow for the bulk component and in green for the surface component. Right panel: Fe 2p$_{3/2}$ core level for Gr/Fe/Ir taken at $h\nu$=800 eV.}
\end{figure}

In Fig.~\ref{fig:C1s} we report the C$1s$ core level for 0.8 ML Fe intercalated between Gr and Ir(111), along with a fit of the different components, performed using Voigt lineshapes (Lorentzian-Gaussian curves taking into account core-hole lifetime and experimental uncertainty, respectively). The C$1s$ exhibits a sharp profiled peak centered at 284.2 eV, associated to the dominant sp$^2$ component of the Gr/Ir system, while two further components are present at higher binding energy, at 284.4 eV and 284.9 eV, due to carbon atoms in valleys and hills of the moir\'e superstructure, respectively. The binding energy of the two extra-peaks is comparable to that of recent experimental results and theoretical prediction of a single Co layer intercalated under Gr/Ir(111), that shows analogous valley-hill corrugation \cite{Pacile_PRB_2014}. In contrast, a recent core level photoemission experiment on graphene CVD-grown directly on the Fe(110) surface \cite{Vinogradov_2017} reports a huge broad C$1s$ component at 284.9 eV, due to a strong interaction of all the C atoms of the graphene layer with the underlying Fe substrate, along with a pronounced shoulder in the photoemission signal at 283.3 eV, suggesting the presence of iron carbide in the sample or at least carbon atoms dissolved in the iron substrate \cite{Vinogradov_2017}. It is worth noting that in our Gr/Fe/Ir(111) hetero-structure, not any Fe-C component due to the possible formation of Fe-carbides is detectable. 

In the left panel of Fig.~\ref{fig:Fe2p}, we report the Ir $4f_{7/2}$ core level of the Gr/Ir(111) substrate before and after intercalation of Fe. In the Gr/Ir(111) system, the neat surface core level component of this Ir surface is maintained, unperturbed by the graphene layer \cite{Scardamaglia_JPCC_2013}. On the other hand, the reduction of the Ir surface state after 0.8 ML Fe intercalation, is due to the  Fe-iridium interaction, as observed also when a single layer of Co or FeCo are intercalated~\cite{Avvisati_JPCC_2017,Avvisati_ASS_2020}. The Fe 2p$_{3/2}$  core level reported in the right panel of Fig.~\ref{fig:Fe2p} does not present any reacted component due to Fe-C intermixing. Recently, J. Brede et al.~\cite{Brede2016} have observed an energy shift of the Ir 4f core levels by about 150 meV towards lower binding energies due to Fe interdiffusion in the Ir substrate, forming a superficial alloy, already at 600-650 K. Keeping the Fe intercalation temperature at 500K, thus avoiding alloy formation, we do not appreciate any energy shift in the photoemission of Ir core levels.

\section{LEED spot intensity analysis}
\begin{figure}
\centering
\includegraphics[clip,width=0.5\textwidth]{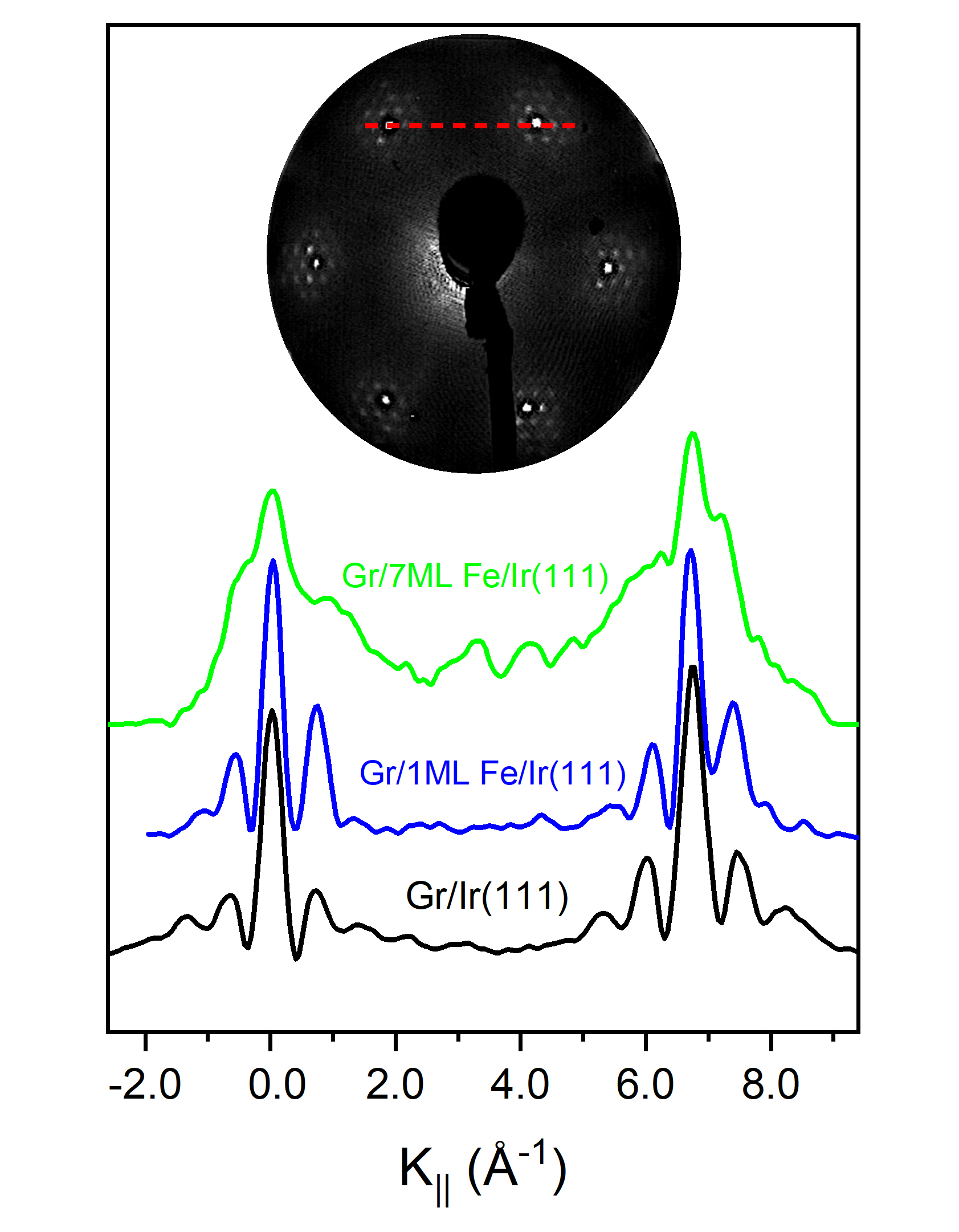}
\caption{\label{fig:LEED} Top: LEED diffraction pattern for Gr/Ir(111) taken at 140 eV. Diffraction spot intensity profile of LEED patterns taken along azimuthal direction [$\overline{1}\overline{1}$2] (red dashed line in LEED patten) from Gr/Ir(111), Gr/1ML Fe/Ir(111) and Gr/7ML Fe/Ir(111) (from bottom to top); the corresponding LEED patterns are reported in Figure 2 of the main paper. } 
\end{figure} 

The spot intensity profiles of the LEED patterns  for Gr/Ir (111), Gr/1ML Fe/Ir(111), and Gr/7ML Fe/Ir(111) taken at 140 eV primary energy are reported in Fig.~\ref{fig:LEED}. The spot intensity profile was taken along the [$\overline{1}\overline{1}$2] azimuthal direction (path shown in the LEED pattern). The spot profiles for Gr/Ir(111) and Gr/1 ML Fe/Ir(111) along with the moir\'e pattern peak intensities are comparable, confirming that the Fe intercalated layer is commensurate to the Ir(111) surface lattice parameter. After further Fe intercalated layers the spot profile of Gr/7ML Fe/Ir(111) unveils broader peaks with a decreased intensity of the extra-spots and a mismatch by about 8\% of the main peaks along the [$\overline{1}\overline{1}$2] direction, with respect to the Ir(111) one, but it is not commensurate to the graphene covering layer. The LEED pattern does not present any longitudinal stripes while a slight signal due to a weakened moir\'e pattern is still distinguishable for all the spots observed in the main diffraction directions, as also the hexagonal symmetry is preserved overall the sample.
The absence of longitudinal pattern and the periodicity of the spots cannot be associated to the structural configuration proposed for graphene directly grown on  bcc-Fe(110) where the distorted hexagonal symmetry reveals an array of 1D stripes with a wave-like longitudinal pattern with periodicity of 17~\AA\ along the stripes \cite{Vinogradov_2012}. In conclusion, the Fe intercalated multilayers present a strained structural arrangement with hexagonal symmetry slightly mismatched with respect to the graphene topmost layer.

\section{Projected Bands and Density of States}

\begin{figure*}
\centering
\includegraphics[clip,width=0.70\textwidth]{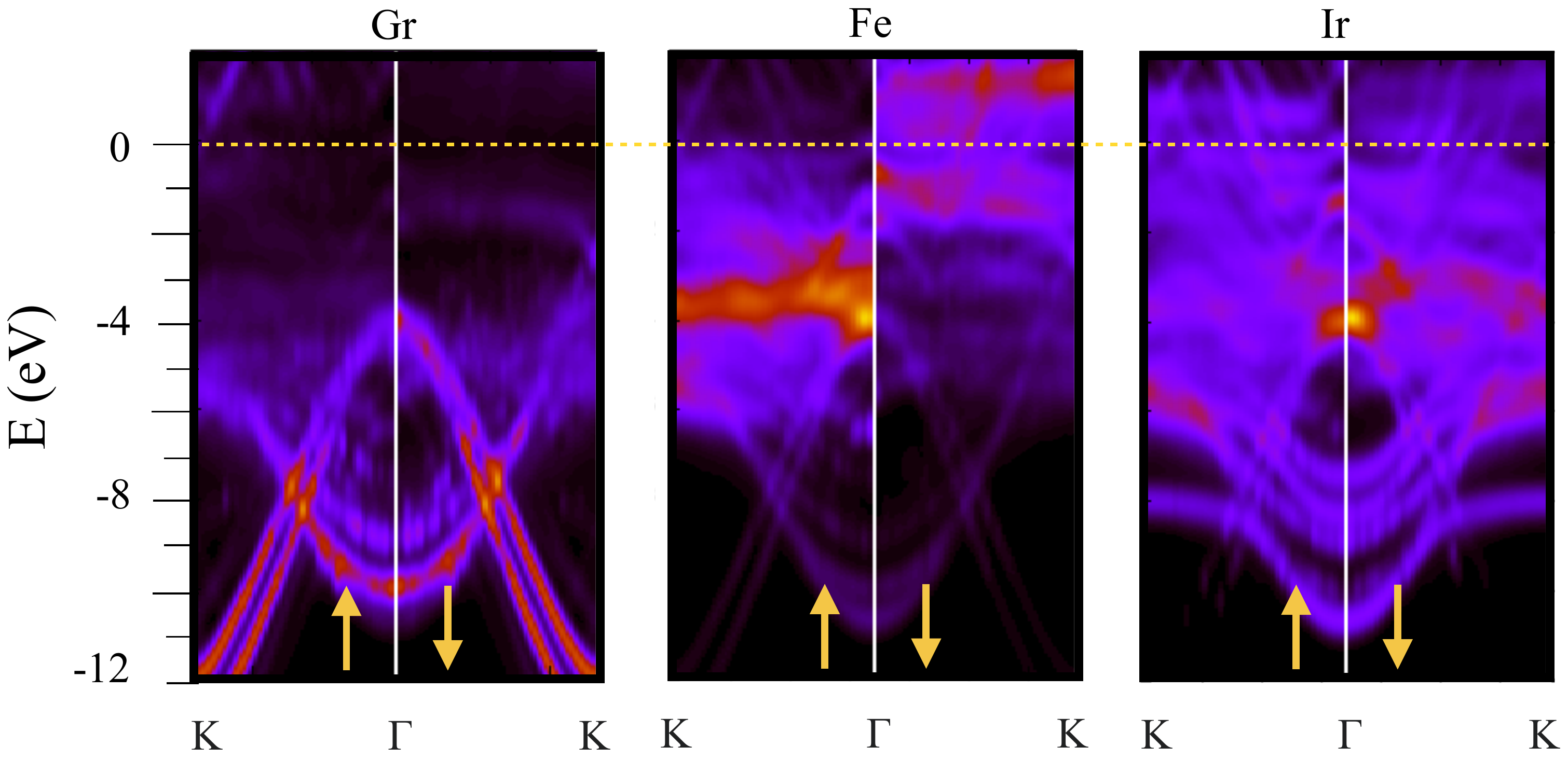}
\caption{\label{fig:pbands_GrFeIr} Bands computed for the Gr/1ML-Fe/Ir(111) unfolded on the Graphene 1$\times$1 cell and projected on C (graphene), Fe, and Ir.}
\end{figure*} 
\begin{figure*}
\centering
\includegraphics[clip,width=0.80\textwidth]{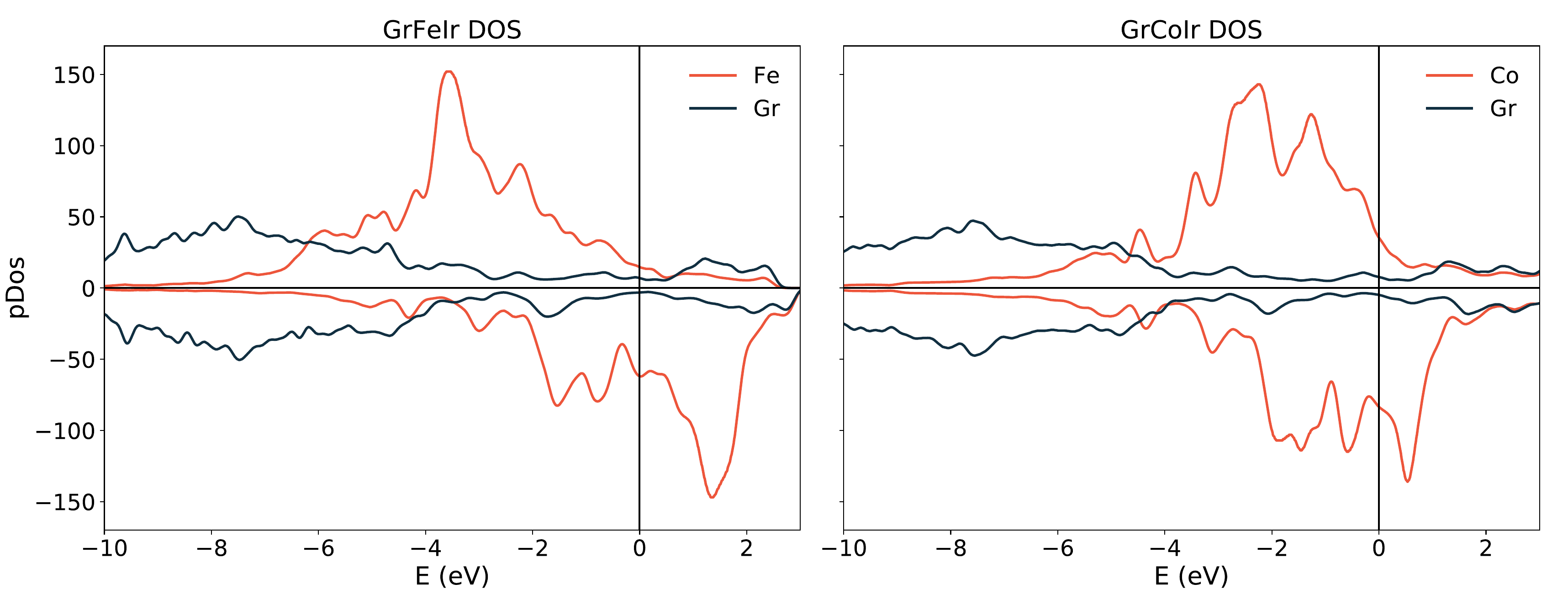}
\caption{\label{fig:pDOS_FeCo} Comparison of the Density of states projected on graphene and Fe or Co, computed for the Gr/1ML-Fe/Ir(111) and Gr/1ML-Co/Ir(111)}
\end{figure*}

In order to better compare with Gr/Ir and to support the identification and assignment of the electronic structure features described in Sec.~III B of the main paper, in Fig.~\ref{fig:pbands_GrFeIr} we plot the unfolded bands of GrFeIr, computed analogously to those in Fig.~4 (main text), but projected on C, Fe, and Ir atoms.
In particular, this allows one to better appreciate the downshift of graphene-related states, the position of Fe-related states in the majority and minority channels, and the role of Iridium states.

In order to compare with the case of GrCoIr,
in Fig.~\ref{fig:pDOS_FeCo} we present the Density of States, projected onto C, Fe and Co orbitals, computed within DFT for Gr/Fe/Ir(111) and Gr/Co/Ir(111), including the complete moir\'e induced periodicity by using a 9$\times$9 supercell of Ir(111), corresponding to a 10$\times$10 supercell of pristine Gr. 

As discussed in the main text, LDA+U with U=2 is used for GrFeIr, while U is not included for GrCoIr.
The main differences between the two systems are seen in the Fe and Co pDOs. The first presents a larger spin split, with the majority $d$ states almost fully occupied, and less occupied minority states, when compered with Co. Fe also shows a pronounced peak around 4~eV below the Fermi energy. These Fe $d$ states are resonant hybridized with the graphene $\pi$ states.